
\magnification=\magstep1
\parskip 0pt
\parindent 15pt
\baselineskip 16pt
\hsize 5.53 truein
\vsize 8.5 truein

\font\titolo = cmbx9 scaled \magstep2
\font\autori = cmsl9 scaled \magstep2
\font\address = cmr9
\font\abstract = cmr8

\hrule height 0pt
\rightline{POLFIS-TH.02/93}
\vfill
\centerline{\titolo CLUSTER VARIATION APPROACH}
\vskip .05truein
\centerline{\titolo TO THE RANDOM--ANISOTROPY}
\vskip .05truein
\centerline{\titolo BLUME--EMERY--GRIFFITHS MODEL}
\vskip .15truein
\centerline{\autori Carla Buzano$^a$, Amos Maritan$^b$ and Alessandro
Pelizzola$^a$}
\vskip .15truein
\centerline{\abstract
$^a$Dipartimento di Fisica and Unit\'a INFM,Politecnico di Torino,
I-10129 Torino, Italy}
\centerline{\abstract
$^b$Dipartimento di Fisica, Universit\'a di Padova, and}
\centerline{\abstract
INFN Sez. di Padova, I-35131 Padova, Italy}

\vfill

\centerline{\bf ABSTRACT}
\vskip .05truein {\address The random--anisotropy Blume--Emery--Griffiths
model, which has been proposed to describe the critical behavior of
$^3$He--$^4$He mixtures in a porous medium, is studied in the pair
approximation of
the cluster variation method extended to disordered systems.
Several new features, with respect to mean
field theory, are found, including a rich ground state,
a nonzero percolation threshold, a reentrant coexistence curve and a
miscibility gap on the high $^3$He concentration side down to zero
temperature.
Furthermore, nearest neighbor correlations are introduced in the random
distribution of the anisotropy, which are shown to be responsible for the
raising of the critical temperature with respect to the pure and
uncorrelated random cases and contribute to the detachment of the
coexistence curve from the $\lambda$ line.}
\vskip .35 truein
\vfill
\eject

\parindent 15pt

{\bf 1. Introduction} \par
\medskip
The study of the effects of randomness on phase transitions has a long
story and only recently there has been considerable experimental and
theoretical efforts to deeply understand them.
Here we will be concerned with a simple model for the phase separation
of $^3$He--$^4$He mixtures in aerogel[1]. \par
Aerogel is an extremely porous medium, the silica glass, made via the
sol--gel process[2]. Its porosity may be as high as $99.8\%$[3]
corresponding to a density of $4 {\rm mg}/{\rm cm}^3$. There is
experimental evidence[2,4] that the aerogel microstructure is rather
ramified, composed of silica strands with a thickness of order $10$\AA \ and
with an area of order $700 {\rm m}^2/{\rm gr}$[5]. \par
In order to describe in a simple way the effects of the porous medium on
the phase separation diagram of $^3$He--$^4$He we model the aerogel as a
random external field which selects which of the two types of Helium to
have nearby. Thus it is quite interesting, among various other things, to
understand the role of the two types of impurities, i.e. the annealed
$^3$He and the quenched randomness of the external field,
on the superfluid transition of $^4$He. \par
Using mean field theory, transfer matrix and real space renormalization
group calculations it was possible to predict a variety of physically
acceptable scenarios for the phase separation diagram as the density of
the quenched impurities is varied[1].
The main features are the disappearance of the
tricritical point present only in the pure $^3$He--$^4$He system[6] and
superfluidity in two coexisting phases, one rich in $^3$He and the other in
$^4$He. Recent experiments[7] have confirmed the above qualitative picture.
\par
The goal of this paper is to give a detailed study of the model proposed in
[1] using the pair approximation of the cluster variation method (CVM)[8],
generalized to random systems[9,10]. This approximation improves the standard
mean field by taking into account effects of correlation among spins at
different sites. Most of the qualitative features of the mean field phase
diagram are preserved. \par
Our approach allows us to incorporate effects of nearest neighbors (n.n.)
correlations in
the quenched impurities, which are essential to a full understanding of the
role of aerogel in the $^3$He--$^4$He phase diagram. Unlike the coexistence
curve, the $\lambda$ line separating the
normal fluid from the superfluid, is rather sensitive to correlations of the
quenched impurities at high $^3$He concentration.
Interesting new
features are found in the $T = 0$ phase diagram and the role of
the percolation threshold is elucidated. \par
In the next section the random--anisotropy Blume--Emery--Griffiths model
(RABEG) will be introduced. The CVM generalized to random systems
will be discussed in detail in connection with the RABEG model. Sec. 3 is
devoted to the discussion of the phase diagram obtained in the present
approximation, also in connection with experimental results, and in
Sec. 4 some conclusions are drawn. \par\bigskip

{\bf 2. Definition of the model and the CVM pair approximation} \par
\medskip
The bulk phase diagram of $^3$He--$^4$He is well described by the
Blume--Emery--Griffiths (BEG) model[6] which is a spin--1 system with
nearest neighbor interactions and a uniform anisotropy external
field distinguishing only between the $\pm 1$ and $0$ values of the spin.
Its generalization to the $^3$He--$^4$He mixture in a porous medium is
described by the hamiltonian[1]
$$H = - J \sum_{\langle i j \rangle}S_i S_j - K \sum_{\langle i j \rangle}
S_i^2 S_j^2 + \sum_i \Delta_i S_i^2 ,\eqno(1)$$
where the spin variables $S_i = 0, \pm 1$ are defined on sites of a lattice
$\Lambda$. The first and the second sums
are over n.n. pairs. When dealing with $^3$He--$^4$He
mixtures[6] $S_i = \pm 1$ and $S_i = 0$ represents $^4$He and $^3$He
respectively, and the superfluid transition is associated to the
spontaneous symmetry breaking between the $\pm 1$ states. The exchange
interaction with strength $J > 0$ is responsible for the superfluid ordering,
while the random anisotropy $\Delta_i$ is related to the difference of chemical
potentials $\mu_3
- \mu_4$  at site $i$, and $K = K_{33} + K_{44} - 2K_{34}$, where
$K_{\alpha \beta}$ is the interaction energy between
$^\alpha$He--$^\beta$He atoms. $K_{\alpha \beta}$ does not depend
significantly on $\alpha$ and $\beta$ and then it is generally assumed $K = 0$.
\par
Since we are going to study a model for $^3$He--$^4$He mixtures in aerogel,
we will let each $\Delta_i$ take the value $\Delta_0$ or $\Delta_1$ with the
same meaning as in [1], i.e. $\Delta_0 < 0$ is the value of the anisotropy
at the pore--grain interface of aerogel (where $^4$He prefers to stay), while
$\Delta_1$ can be thought of as a bulk field which controls the $^3$He
concentration $x$. \par
Two order parameters can be defined: $m = \langle \langle S_i \rangle
\rangle_\Delta$ and $q = \langle \langle S_i^2 \rangle \rangle_\Delta$,
which are the quenched averages of the thermal equilibrium values $\langle
S_i \rangle$ and $\langle S_i^2 \rangle$. $m$ is the superfluid order
parameter, while $q = 1 - x$ is the $^4$He concentration. \par
As we mentioned in the introduction we take
into account n.n. correlations in the random field
distribution assuming the following joint probability density (with $i$ and
$j$ nearest neighbors) for the $\Delta_i$'s:
$$P(\Delta_i,\Delta_j) = \sum_{n_1,n_2=0,1} p_{n_1 n_2}
\delta(\Delta_i - \Delta_{n_1}) \delta(\Delta_j - \Delta_{n_2}),\eqno(2)$$
$$p_{n_1 n_2} = p_{n_1} p_{n_2} + (2 \delta_{n_1 n_2} - 1)\epsilon p^2,
\eqno(3)$$
where $p_0 = p$ is the fraction of sites at the aerogel interface (the
grain space is neglected), $p_1 = 1 - p$ and $\epsilon$
is a measure of the interface correlation. Integrating
$\Delta_j$ one obtains
$$P(\Delta_i) = p\delta(\Delta_i - \Delta_0) + (1 - p)\delta(\Delta_i -
\Delta_1), \eqno(4)$$
which was used as a starting point in [1]. \par
We are now ready to build the pair CVM free energy following the procedure
outlined by Morita[9] for a general random system, which we briefly review.
\par
Consider a random system on a finite lattice, for which the configuration of
random fields and
interactions is specified by a unique random variable $h$ with
distribution $P(h)$, and define, according to the usual rules of statistical
mechanics, a density matrix $\rho_c(\sigma \vert h)$ for each configuration $h$
($\sigma$ stands for the set of statistical degrees of freedom of the system,
e.g. spin variables). In terms of $\rho_c$, the free energy associated to
the configuration $h$ will be given by
$$F(h) = \sum_\sigma \rho_c(\sigma \vert h) \left[ H(h,\sigma) + k_B T \ln
\rho_c(\sigma \vert h)
\right], \eqno(5)$$
where $H(h,\sigma)$ is the hamiltonian, $k_B$ Boltzmann's constant, and $T$
absolute
temperature. The quenched free energy $F$ will then be given by
$$F = \sum_h P(h) F(h). \eqno(6)$$
Upon introducing the generalized density matrix
$$\rho(h,\sigma) = P(h) \rho_c(\sigma \vert h) \eqno(7)$$
one can easily show that
$$F = \sum_{h,\sigma} \rho(h,\sigma) \left[ H(h,\sigma) + k_B T \ln
\rho(h,\sigma) \right] + T
S_c, \eqno(8)$$
where $S_c$ depends only on $P(h)$ and is given by
$$S_c = - k_B \sum_h P(h) \ln P(h). \eqno(9)$$
As in the pure case, we have a variational principle: $\rho(h,\sigma)$ can be
determined as the matrix which minimizes $F$, as given by (8), with the
constraint
$$\sum_\sigma \rho(h,\sigma) = P(h).\eqno(10)$$
In such a scheme it can be shown that the quenched average of the
expectation value $\langle A(h,\sigma) \rangle$ of an operator $A(h,\sigma)$ is
given by
$$\langle \langle A(h,\sigma) \rangle \rangle_h = \sum_h P(h) \langle
A(h,\sigma)
\rangle = \sum_{h,\sigma} \rho(h,\sigma) A(h,\sigma).\eqno(11)$$ \par
The cluster variation method can then be obtained by taking the
thermodynamic limit and truncating the cumulant expansion for the entropy
$S = - k_B T \sum_{h,\sigma} \rho(h,\sigma) \ln \rho(h,\sigma)$ to a set of
"maximal preserved
clusters" $\Gamma_i$, $i = 1,2,\ldots r$ (and all their translates).
The variational principle will then be applied to the
reduced density matrices $\rho_{\Gamma_i}(h,\sigma)$ associated to the maximal
preserved clusters. \par
In the following, our maximal preserved
cluster will be the n.n. pair, and thus we will introduce a density matrix
$\rho_p$
with elements $\rho_p(h,\sigma) \equiv \rho_p^{(n_1,n_2)}(S_1,S_2)$
($n_1, n_2 = 0,1$, $S_1, S_2 = \pm 1,0$), where $h =
(\Delta_{n_1},\Delta_{n_2})$ is the random field configuration on the cluster
and $\sigma = (S_1,S_2)$ denotes the set of spin variables. $\rho_p$ (which is
diagonal, since we are dealing only with the $z$-component of the spins)
is subject to the constraint $\sum_\sigma \rho_p(h,\sigma) = P(h)$, that is
$$\sum_{S_1 S_2} \rho_p^{(n_1,n_2)}(S_1,S_2) = p_{n_1 n_2}.\eqno(12)$$
Writing $\rho_p$ as a direct sum of density
matrices, one for each possible configuration of the random fields on the
cluster:
$$\rho_p = \bigoplus_{n_1,n_2 = 0,1} p_{n_1 n_2} \tilde\rho_p^{\ (n_1,n_2)}
\eqno(13)$$
the constraint (12) becomes
$${\rm Tr} \tilde\rho_p^{\ (n_1,n_2)} = 1, \qquad n_1,n_2 = 0,1. \eqno(14)$$
Furthermore, $\rho_p$ has the obvious symmetry property
$$\rho_p^{(n_1,n_2)}(S_1,S_2) = \rho_p^{(n_2,n_1)}(S_2,S_1) \ge 0.\eqno(15)$$
As usual, a (reduced) site density matrix can be obtained from $\rho_p$ by
a partial trace:
$$\rho_s = \matrix{ \phantom{^{p \backslash s}} \cr {\rm Tr} \cr
^{p \backslash s} \cr} \rho_p, \eqno(16)$$
or, more explicitly,
$$\rho_s^{(n_1)}(S_1) = \sum_{n_2 S_2} \rho_p^{(n_1,n_2)}(S_1,S_2).\eqno(17)$$
\par
The pair CVM free energy is then given by
$$F = U + k_B T \left[ (1 - \nu) {\rm Tr} (\rho_s \ln \rho_s) + {\nu \over 2}
{\rm Tr} (\rho_p \ln \rho_p) \right],\eqno(18)$$
with $U = \langle
\langle H \rangle \rangle_\Delta$ and $\nu$ coordination number of the
lattice, and has to be minimized with respect to
$\rho_p$, with the constraints (14) and (15). Taking derivatives of $F$ with
respect to $\rho_p$ leads to three systems of equations, which are linear in
the
elements of $\ln \tilde\rho_p^{\ (0 0)}$, $\ln \tilde\rho_p^{\ (0 1)}$ and
$\ln \tilde\rho_p^{\ (1 1)}$ respectively, and where the elements of
$\ln \rho_s$ can be treated as parameters. Exponentiating the solution and
making
use of (14) yields
$$\rho_p^{(n_1,n_2)}(S_1,S_2) = p_{n_1 n_2}
{e^{{\beta \over \nu} S_1 S_2}
V_{S_1}^{(n_1)} V_{S_2}^{(n_2)} \over {\displaystyle \sum_{S_1,S_2}}
e^{{\beta \over \nu} S_1 S_2}
V_{S_1}^{(n_1)} V_{S_2}^{(n_2)}},\eqno(19)$$
with $\beta = \nu J / k_B T$ and
$$V_\pm^{(n)} = e^{-\displaystyle{\beta \over \nu} d_n}
\left[ q_n \pm m_n \over 2 (1 - q_n) \right]^\alpha, \qquad
V_0^{(n)} = 1,$$
where $\alpha = (\nu - 1)/\nu$, $d_n = \Delta_n/(\nu J)$.
The parameters $m_n$ and $q_n$, which can be thought of as "local order
parameters" at the aerogel interface ($n = 0$) and in the interior ($n =
1$), are given by
$$\eqalign{& m_n = \tilde\rho_s^{(n)}(+) - \tilde\rho_s^{(n)}(-) \cr
& q_n = \tilde\rho_s^{(n)}(+) + \tilde\rho_s^{(n)}(-) \cr}.\eqno(20)$$
It can be easily shown that the parameters
$m_n$ and $q_n$ are solutions of the following equations
$$\left\{\eqalign{& p_n m_n = \sum_{n_2,S_2} \left(
\rho_p^{(n,n_2)}(+,S_2) - \rho_p^{(n,n_2)}(-,S_2) \right) \cr
& p_n q_n = \sum_{n_2,S_2} \left(
\rho_p^{(n,n_2)}(+,S_2) + \rho_p^{(n,n_2)}(-,S_2) \right) \cr}\right.
\eqno(21)$$
which, together with (19), are the basic results of this
section, and the starting point for the analysis of the phase diagram of
our model at finite temperature. \par
Finally, the order parameters of the model are related to our "local order
parameters" by
$$\eqalign{ & m = p m_0 + (1 - p) m_1 \cr
& q = p q_0 + (1 - p) q_1 \cr}. \eqno(22)$$
\vfill\eject
{\bf 3. The phase diagram} \par
\medskip
At zero temperature, several different phases can be found, which will be
identified by the set $(m_{00} = q_{00}, m_{01} = q_{01}, m_{10} = q_{10},
m_{11} = q_{11})$, where
$$\eqalign{
& m_{n_1 n_2} = \sum_S \left( \tilde\rho_p^{(n_1,n_2)}(+,S) -
\tilde\rho_p^{(n_1,n_2)}(-,S) \right) \cr
& q_{n_1 n_2} = \sum_S \left( \tilde\rho_p^{(n_1,n_2)}(+,S) +
\tilde\rho_p^{(n_1,n_2)}(-,S) \right) \cr}. \eqno(23)$$
These are related to the expectation values of $S_i$ and $S_i^2$
respectively when the anisotropy field is $\Delta_{n_1}$ at site $i$ and
$\Delta_{n_2}$ in one of its neighbours.\par
The phase diagram is
easily determined by comparing the energies $U$
of the different phases, and is reported in Fig. 1. Two comments are in
order: i) as opposite to the mean field result[1], the boundaries between
the different ground states do no
longer depend on the probability distribution of the random variables
and ii) increasing the level of the approximation we have obtained a more
complicated phase diagram, with a larger number of possible ground states.
This is reminescent of what happens in the exact solution at $T=0$
of the random field Ising model on a Bethe lattice[11]. Notice
that our model (1) reduces to a random field Ising model for $J=0$
and $K \ne 0$ since all spin variables appear squared and thus assuming
only the two values $0,1$.\par
At finite temperature, and for $\Delta_0 < 0$,
it has been shown in the mean field analysis[1] that
the phase diagram has a quite rich structure, with a second order
transition separating the ordered phases from the disordered
one, a first order transition, terminating in a critical point,
between the ordered phases above, and several multicritical points.
Our approximation yields an even richer structure: indeed we have three
different ground states (see dotted line in Fig. 1), $(1111)$ for $d_1 <
1/2$ corresponding to $^4$He present everywhere, $(1110)$ for
$1/2 < d_1 < 1$, which should correspond to $^4$He present in the aerogel
interface and diffusing near it,
and $(1100)$ for $d_1 > 1$, i.e. $^4$He only at the aerogel
interface. Again,
ordered phases are separated by the disordered (high temperature) one by a
second order transition line, for which an equation can be obtained
by expanding (23) around $m_0 = m_1 = 0$. Up to the
first order, the equations for $m_n$ yield
$$\left\{ \eqalign{ & m_0 = a_{00} m_0 + a_{01} m_1 \cr
& m_1 = a_{10} m_1 + a_{11} m_1 \cr} \right. , \eqno(24)$$
where
$$a_{\gamma \lambda} = \alpha \left[ \delta_{\gamma \lambda} + 4 {p_{\gamma
\lambda} \over p_\gamma q_\lambda} {V^{(\gamma)} V^{(\lambda)} \over
Z_{\gamma \lambda}} \ {\rm sinh} {\beta \over \nu} \right]. \eqno(25)$$
The equation for the critical temperature turns out to be
$$(a_{00} - 1)(a_{11} - 1) = a_{01} a_{10}, \eqno(26)$$
meaning that a  non trivial solution of eq. (24) exists,
where $q_0$ and $q_1$ must be determined from the corresponding equations
with $m_0 = m_1 = 0$. A solution for the critical temperature is obtained
for any $d_1$ only if $p$ is greater than the percolation threshold $p_c$. \par
The percolation threshold can be defined as
follows: in the limit $\Delta_1 \to +\infty$, $S_i$ will be $0$ for all sites
with anisotropy $\Delta_1$ and our model will be equivalent to a
random--diluted BEG model with anisotropy $\Delta_0$ and concentration $p$.
Such a
model will undergo a second order transition at $T_c > 0$ for any large
enough $p$, and the percolation threshold is just that value of $p$ for
which $T_c$ becomes zero. It can be obtained by taking the limits $\Delta_1
\to +\infty$ and $\beta \to +\infty$ in the equation for the critical
temperature, and the result is
$$p_c = {1 \over (\nu - 1)(1 + \epsilon)},\eqno(27)$$
representing a remarkable improvement with respect to the mean field
theory, which gives $p_c = 0$. Notice that for uncorrelated
disorder, i.e. $\epsilon=0$, the exact result for the Bethe lattice
with coordination $\nu$ is recovered[12].\par
For $p > p_c$ the phase diagram has the structure shown in Figs. 2(a) and
2(b) (heavy lines):
solid lines are used for the second order transition, while dashed lines
stand for the first order transition separating $(1111)$ and $(1110)$
phases (if in the $(d_1,\tau = \beta^{-1})$ plane) or for the
corresponding coexistence
region (if in the $(x,\tau)$ plane, where $x$ is the $^3$He concentration).
$C$ is a critical point. No first
order line separating $(1110)$ and $(1100)$ phases seems to
start from the point $F(d_1 = 1, \tau = 0)$, where a first order transition
occurs at zero temperature. Again this is reminescent of the behaviour
of the random field Ising model on the Bethe lattice[11].
The phase diagram at $T > 0$ is qualitatively the
same as that given by mean field theory, which is reported for comparison
(light lines). In all considered cases ($\Delta_0<0$) $x<1-p$. \par
For $0 < p < p_c$ a new feature arises, that cannot be described by mean
field theory: the second order line has a
limit point $F$ at zero temperature and $d_1 = 1$, independent of $p$,
$\epsilon$ and $d_0$, and the critical temperature is zero for any $d_1 > 1$,
meaning that, even if the ground state is ordered, aerogel
concentration is too low for the pore--grain
interface to sustain an ordered phase at finite (i.e.
nonzero) temperature.
The corresponding phase diagrams are given in Figs. 3(a) and 3(b)
and in Figs. 4(a) and 4(b), for typical values of the parameters.
In the latter case the second order
line is reentrant and intersects the first order line in the critical end
points $A$ and $B$. It is easily realized that there exists a value $p^*
< p_c$ of $p$ for which the points $A$ and $B$ merge into one, giving rise
to a new multicritical point where the second and the first order lines are
tangent with respect to each other. It is also interesting to remark that
the first order transition has always a reentrant behavior, a feature
that does not exist in the mean field treatment[1]. \par
In our scheme the behavior of order parameters and n.n. correlation
functions can be obtained in a natural way: in Fig. 5 we report the behavior
of the local order parameters $m_n$ and $q_n$ vs. temperature for the case
$1/2 < d_1 < 1$ (dotted line in Fig. 3(a)),
where $m_0 = q_0 = 1$ and $m_1 = q_1 = p_{10}/(1 - p)$
in the ground state. It is interesting that $m_0$ and $m_1$
exhibit a quite unusual oscillation. Furthermore, in Fig. 6,
we report the global order parameter
$m$ vs. temperature at fixed concentration, corresponding to the dotted
line in Fig. 2(b). \par
Finally, let us discuss the effect of a correlation $\epsilon > 0$ in the
random distribution on the phase diagram, and especially on the second order
transition. In the mean field analysis[1] it was shown that the second order
transition occurs at $\tau = 1 - x = q$, independent of $p$, while our analysis
shows that the second order critical temperature at fixed concentration
is indeed weakly dependent
on $p$ (Fig. 7). Introducing a correlation $\epsilon$ of order unity in the
random distribution of the $\Delta_i$'s causes the critical temperature to
raise significantly, especially in the low--$q$ (high--$x$) region (Fig. 8),
and we believe that this
effect should account for the experimentally observed[7] increase in the
critical
temperature with respect to the pure case. The behavior of $T_c$ as
a function of $\epsilon$ is shown in Fig. 9. \par \bigskip
{\bf 4. Conclusions} \par \medskip
A model for the critical behavior of $^3$He--$^4$He mixtures in a porous
medium (aerogel), namely the random--anisotropy Blume--Emery--Griffiths
model[1], has
been investigated in the pair approximation of the cluster variation
method. This improves the mean field analysis given in [1] in several
directions. A rather rich phase diagram is found already at zero
temperature, with various ground states corresponding to ($\Delta_0 < 0$):
$^4$He present
everywhere, (1111) in Fig. 1, at the aerogel interface and in a neighbour
of it (1110), and only at the aerogel interface (1100). The symmetric
situations, with $^3$He preferentially near aerogel are of course possible
within the context of a model (Fig. 1 for $d_0 > 1/2$). \par
A nonzero percolation threshold can be
determined, above which the phase diagram is mean field like, and below
which it changes qualitatively, with the appearance of a zero temperature
limit point for the second order transition line.
A completely open and extremely interesting question concerns the
universality class of the whole line of continuous transitions: is
there a zero temperature or $d_1 = + \infty$ fixed point attracting a
finite part of the critical line and an unstable fixed point at finite
temperature separating two different critical behaviors (e.g. $d_1 < 0$ and
$d_1 \approx 1$)? A real space renormalization group analysis might be
suitable to address this problem. \par
In all considered cases, the first
order transition line (in the temperature--anisotropy phase
diagram) and the coexistence region (in the temperature--concentration
phase diagram) exhibit reentrant behavior.
This reentrance is also partially responsible for the existence of a
miscibility gap on the high $^3$He concentration side going down to zero
temperature. \par
The experimental phase diagrams are in remarkable agreement with the ones
presented in figs. 2(b) and 3(b). Quantitative comparisons would of course
require more adequate treatments of the superfluid transition. \par
Particular attention has been devoted to the analysis of the effects of a
nearest neighbor correlation in the random distribution of the anisotropy,
which should not be neglected due to the high correlation characterizing
the aerogel interface. The main result is the increase in the critical
temperature with respect to the pure case, contributing to the detachment
of the $\lambda$ line from the coexistence curve and which has been
experimentally observed [7]
and cannot be explained in terms of the uncorrelated
randomness alone. Correlations in the quenched impurities make the question
of the universality class of the critical line more involved even at low
disorder concentration. \par \bigskip
{\bf Acknowledgements} \par \medskip
We are indebted to Jayanth Banavar, Moses Chan and Flavio Toigo for enlighting
discussions. We also thank Moses Chan for sending us the results of ref.
[7] before publication. \par
\vfill\eject
{\bf References} \par
\medskip
\item{[1]} A. Maritan, M. Cieplak, M.R. Swift, F. Toigo and J.R. Banavar,
Phys. Rev. Lett. {\bf 69} (1992) 221.
\item{[2]} J. Fricke, {\it Aerogel} in Scientific American {\bf 258} (1988) 92.
\item{[3]} T.M. Tillotson, L.W. Hrubesh and I.M. Thomas, {\it Partially
Hydrolized Alkoxysilanes as Precursors of Silicon Aerogels}, in Better
Ceramic through Chemistry III, C. Brinker, D. Clark and D. Ulrich ed., Mat.
Res. Soc. Proc. {\bf 121} (1988) 685.
\item{[4]} D.W. Schaefer, C.J. Brinker, D. Richter, B. Farago and B. Frick,
Phys. Rev. Lett. {\bf 64} (1990) 2316.
\item{[5]} A. Wong, {\it Liquid--Vapor Critical Point of Fluids in Porous
Glasses}, Ph.D. thesis, Penn State University (1991).
\item{[6]} M. Blume, V.J. Emery and R.B. Griffiths, Phys. Rev. {\bf A 4}
(1971) 1071.
\item{[7]} J. Ma, S. Kim and M.H.W. Chan, in preparation.
\item{[8]} R. Kikuchi, Phys. Rev. {\bf 81} (1951) 988; G. An, J. Stat.
Phys. {\bf 52} (1988) 727; T. Morita, J. Stat. Phys. {\bf 59} (1990) 819.
\item{[9]} T. Morita, Progr. Theor. Phys. Suppl. {\bf 80} (1984) 103; see
also H. Falk, J. Phys. {\bf C 9} (1976) L213.
\item{[10]} Ph. de Smedt, J.O. Indekeu and L. Zhang, Physica {\bf A 140}
(1987) 450.
\item{[11]} R. Bruinsma, Phys. Rev. {\bf B 30} (1984) 289.
\item{[12]} M.E. Fisher and J.W. Essam, J. Math. Phys. {\bf 2} (1961) 609.
\vfill\eject

{\bf Figure captions} \par
\parindent .4 truein
\item{}
\itemitem{\hbox to .82 truein{Fig. 1 : \hfill}} Phase diagram at $T = 0$.
\itemitem{\hbox to .82 truein{Fig. 2(a) : \hfill}} Phase diagram in the
$(d_1,\tau)$ plane for $\nu = 6$,
$\epsilon = 0$, $p = 0.21 > p_c$, $d_0 = -0.5$. Heavy lines are our results
and light lines are results from mean field approximation.
\itemitem{\hbox to .82 truein{Fig. 2(b) : \hfill}} Phase diagram in the
$(x,\tau)$ plane for the case of Fig. 2(a). Symbols as in fig. 2(a)
(for the dotted line see Fig. 6).
\itemitem{\hbox to .82 truein{Fig. 3(a) : \hfill}} The same as Fig. 2(a)
with $p = 0.19 < p_c$ (for the dotted line see Fig. 5).
\itemitem{\hbox to .82 truein{Fig. 3(b) : \hfill}} Phase diagram in the
$(x,\tau)$ plane for the case of Fig. 3(a).
\itemitem{\hbox to .82 truein{Fig. 4(a) : \hfill}} The same as Fig. 2(a) with
$p = 0.1 < p_c$.
\itemitem{\hbox to .82 truein{Fig. 4(b) : \hfill}} Phase diagram in the
$(x,\tau)$ plane for the case of Fig. 4(a).
\itemitem{\hbox to .82 truein{Fig. 5 : \hfill}} $m_n$ and $q_n$
vs. $T$ along the dotted line in Fig. 3(a) ($d_1 = 0.65$).
\itemitem{\hbox to .82 truein{Fig. 6 : \hfill}} $m$ vs. $T$ along the
dotted line in Fig. 2(b) ($x = 0.69$).
\itemitem{\hbox to .82 truein{Fig. 7 : \hfill}} The critical temperature in
the $(x,\tau)$ plane for $\nu = 6$, $\epsilon = 0$, $d_0 = - 0.5$ and $p =
0$ (solid line), $0.1$ (dashed line) $0.2$ (dotted line).
\itemitem{\hbox to .82 truein{Fig. 8 : \hfill}} The critical temperature in
the $(x,\tau)$ plane for $\nu = 6$, $d_0 = -0.5$, $p = 0.1$ and $\epsilon =
0$ (solid line), $0.5$ (dotted line), $1.0$ (dot--dashed line) and $2.0$
(dashed line).
\itemitem{\hbox to .82 truein{Fig. 9 : \hfill}} The critical temperature
vs. $\epsilon$ for the case of Fig. 7 and with $x = 0.6$.
\vfill\eject\end